\documentclass[]{aa} 
\usepackage{graphics,amsmath,amssymb}
\def\kms{\relax \ifmmode {\,\mbox{km\,s}}^{-1}\else \,\mbox{km\,s}$^{-1}$\fi}
\def\ha{\relax \ifmmode {\mbox H}\alpha\else H$\alpha$\fi}
\def\hb{\relax \ifmmode {\mbox H}\beta\else H$\beta$\fi}
\def\hi{\relax \ifmmode {\mbox H\,{\scshape i}}\else H\,{\scshape i}\fi}  
\def\hii{\relax \ifmmode {\mbox H\,{\scshape ii}}\else H\,{\scshape ii}\fi}
\def\oiii{\relax \ifmmode {\mbox O\,{\scshape iii}}\else O\,{\scshape iii}\fi}
\def\oii{\relax \ifmmode {\mbox O\,{\scshape ii}}\else O\,{\scshape ii}\fi}
\def\oi{\relax \ifmmode {\mbox O\,{\scshape i}}\else O\,{\scshape i}\fi}
\def\nii{\relax \ifmmode {\mbox N\,{\scshape ii}}\else N\,{\scshape ii}\fi}
\def\sii{\relax \ifmmode {\mbox S\,{\scshape ii}}\else S\,{\scshape ii}\fi}
\def\lha{\relax \ifmmode \mbox {L}_{H\alpha}\else $\mbox{L}_{H\alpha}$\fi}
\def\ldig{\relax \ifmmode {\mbox L}_{DIG}\else ${\mbox L}_{DIG}$\fi}
\def\ls{\relax \ifmmode {\mbox L}_{ Str}\else ${\mbox L}_{ Str}$\fi}
\def\eme{\relax \ifmmode {\,\mbox{pc\,cm}}^{-6}\else \,pc\,cm$^{-6}$\fi}
\def\l{\relax \ifmmode  \lambda\else $\lambda$\fi}
\def\me{$^{-1}$}              

\def\arcsec{\hbox{$^{\prime\prime}$}}
\def\deg{\hbox{$^\circ$}}
\begin{document}

   \thesaurus{11     
              (11.07.1;  
               11.09.1;  
 	       11.09.4;  
               11.19.2;  
               09.08.1)} 

\title{The origin of the ionization of the diffuse ionized gas  in spirals.}

\subtitle{II. Modelling the distribution of ionizing radiation in NGC 157}

\author{A. Zurita\inst{1,}\inst{2}, J. E. Beckman\inst{2,}\inst{3}, M. Rozas\inst{4}
\and   S. Ryder\inst{5}}

   \offprints{A. Zurita (azurita@ing.iac.es)}

   \institute{Isaac Newton Group of telescopes, La Palma, Spain
   \and{Instituto de Astrof\'\i sica de Canarias, C. V\'\i a 
   L\'actea s/n, 38200-La Laguna, Tenerife, Spain} 
   \and{Consejo Superior de Investigaciones Cient\'\i ficas (CSIC), Spain}
   \and{Observatorio Nacional de S. Pedro M\'artir, UNAM, Ensenada, M\'exico}
   \and{Anglo-Australian Observatory, Epping, NSW 1710, Australia}\\ 
  email: azurita@ing.iac.es, jeb@ll.iac.es, maite@astrosen.unam.mx,  
  sdr@aaoepp.aao.GOV.AU}

\date{} \authorrunning{Zurita et al.}  \maketitle

\begin{abstract} 
In this paper we make a quantitative study of the hypothesis that the diffuse \ha\
emitted from the discs of spiral galaxies owes its origin
to the ionizing photons escaping from \hii\ regions. The basis of the models
is the assumption that a fraction of the Lyc luminosity from the OB stars within 
each \hii\ region escapes from the region, leaking into the diffuse gas.
A basic input element of any such model is a position and luminosity catalogue in \ha\
of the \hii\ regions in the galaxy under test, down to a low limiting luminosity,
and we had previously produced a catalogue of this type for NGC~157.
An initial family of models can then be generated in which the Lyc escaping 
from an \hii\ region is parametrized in terms of the observed \ha\ luminosity 
of the region and the escaping fluxes allowed through the diffuse disc gas.
These models can then be refined using a measured map of \hi\ surface density 
to effect the down-convertion of the Lyc to \ha. For NGC~157 an \hi\ map was available.
Although its moderate angular resolution did limit the accuracy 
with which we could test our models, the predicted diffuse \ha\ surface 
brightness distributions from our mo\-dels were
compared with the observed distributions showing that, in general terms, the hypothesis 
of density bounding for the \hii\ regions allows us to predict well the spatial 
distribution of the diffuse ionized gas.
In the model yielding the best fit to the data, the regions of lower luminosity 
lose a constant fraction of their ionizing flux to their surroundings, while for \hii\ 
region luminosities above a specific transition va\-lue the ionizing escape fraction 
is a rising function of the \ha\ luminosity.


\keywords{Galaxies: general--Galaxies: individual (NGC~157)
--Galaxies: ISM--Galaxies: spiral--ISM: general--ISM: H{\scshape ii}
regions.}

\end{abstract}

\section{Introduction}
\label{intro}
Although the  diffuse component of \ha\  emission from a disc  galaxy has low
surface brightness,  the integrated  \ha\ luminosity emitted  by a  galaxy in
this component is re\-la\-ti\-ve\-ly high, of order 50\% of the total radiated by the
galaxy,  for  late-type  spirals  (e.g.  Ferguson et  al.  1996;  Hoopes  et
al. 1996).  In Zurita et al. (2000,  hereinafter Paper~I) we  made a detailed
study of  6 galaxies in  which we quantified  this diffuse component  and its
geometrical  distribution  across  the  face  of each  galaxy.  Supporting  an
observation  previously noted by  Walterbos \& Braun (1994) and Ferguson et  al. (1996),  we found  a very
clear spatial correlation between the  \hii\ region e\-mi\-ssion in \ha\ and that
of the diffuse  emission, leading to a prima facie  case for considering that
OB asso\-cia\-tions  are the sources or,  at least, the principal  sources of the
ioni\-zing photons  which eventually  reach and ionize  the diffuse gas  in the
disc  as  a  whole.  Quantitatively,  it  appears that  the  stars  in  these
associations can  emit su\-ffi\-cient photons  above the Lyman limit  to maintain
 the di\-ffu\-se component ionized, as well as the gas within the \hii\ regions of
the galaxies  observed, and are  an obvious choice  for the primary  cause of
the diffuse \ha. They  are clearly  collectively more  luminous in  the Lyman
continuum (Lyc) that the population of white dwarfs, for example. How\-ever, it
has never been entirely clear that  a major fraction of Lyc photons can escape
from the immediate  su\-rroun\-dings of an OB a\-sso\-cia\-tion:  its \hii\ region, and
even less clear  that an escaping photon can  be transmitted across distances
of  order a  kiloparsec, from  the nearest  OB associations  to zones  in the
diffuse  disc  gas from  which  sig\-ni\-fi\-cant  surface  brightness in  \ha\  is
observed.  This difficulty  was  a key  motive  in leading  Sciama (1990)  to
propose  his  decaying neutrino  model  for  the  ionization of  the  diffuse
com\-po\-nent, although recent  EUV observations of the interstellar medium (ISM)
  close to the  Sun (Bowyer et al.
1999) appear to have ruled out this model.

In  the case  of  the Galaxy,  several  models have  been  proposed in  which
transport of  ionizing photons over distances of  order 1 kpc is  shown to be
viable, at least in the neighbourhood of the Sun (Miller \& Cox 1993; Dove \&
Shull 1994). The fact that locally we  can quantify the OB stars, and can map
the ionized and neutral hydrogen in detail, means that models of this sort are
useful since they can be well compared with data. 

There are two main physical
reasons why ionizing photons can escape from \hii\ regions and transit large
distances in spite of the large ionization cross--section of  neutral H. The
principal reason is the inhomogeneity of the ISM. It is
well established, for exam\-ple, that although the r.m.s. electron density in a
luminous \hii\ region is of order 1-5 cm$^{-3}$ (e.g. Kennicutt 1984; Rozas
et al. 1996b, 1999), the electron density determined locally via emission line
ratios  is of  order  100 cm$^{-3}$  (e.g.  Zaritsky et  al.  1994). This  is
interpreted (Osterbrock \& Flather 1959; Osterbrock 1989)  as reflecting the
presence of  dense knots (or
clumps of gas)  which are the sites of virtually all the optical line emission
from  the  \hii\ region,  within  an ambient  volume  of  much lower  density
gas. The ratio of  the volume of dense gas to the  total volume of the region
is embodied  in the  ``filling factor",  which may be  defined simply  as the
ratio of the  two volumes. This rather simple scheme of dense clumps 
in a sparse medium (i.e. the inhomogeneous structure of the ISM) is
backed  by the  general theory  of  thermal equilibrium  in the  interstellar
medium,  treated  classically  by   Field,  Goldsmith  \&  Habing  (1969):  a
two--phase model,  with a  warm substrate at  T$\sim$10000 K, and  cool dense
clumps  at T$<$100  K,  then by  Cox \&  Smith  (1974): a  three phase  model
including also  the effects of supernovae  and a component  at T$>$100000 K,
and McKee \& Ostriker (1977), who produced a more complete three--phase model
against which much observational  material within the Galaxy has subsequently
been compared. It can be demonstrated (see e.g. Spitzer 1978) that in the ISM
distinct  phases should  co-e\-xist in  pressure equilibrium,  while  states of
intermediate  density and  temperature are  unstable and  decay  rapidly. The
scenario of dense clouds  with $N_{e}\sim$100~cm$^{-3}$, surrounded by a much
lower  density substrate, with  $N_{e}\lesssim$ 1~cm$^{-3}$,  and infiltrated
with high  temperature fully ionized  plasma having $N_{e}<10^{-2}$~cm$^{-3}$
has  been  well   verified  by  many  detailed  measurements   in  the  local
interstellar medium (LISM) within a few hundred parsec of the sun (Trapero et
al. 1995, 1996). The
high den\-si\-ty  clo\-uds con\-tribute a  major fraction of  the mass, but  occupy a
minor fraction of  the volume (Trapero et al. 1996). Stu\-dies by Berkhuijsen
(1999) in  the Galactic  ISM show that  this type of  inhomogeneous structure
prevails over a wide range of mean ISM density and temperature. 
There is evidence for this structure also in external galaxies, as shown by 
Braun (1997) for a sample of the 11 nearest spiral galaxies beyond the Local 
Group.
The fact that
an inhomogeneous  ISM offers a lower effective  volume absorption coefficient
for Lyc photons than a homogeneous  medium is easy to demonstrate, and a very
general result, common to many types of transfer problems.

The second effect permitting a fraction  of Lyc photons to have free paths on
kpc scales is a secondary consequence of the inhomogeneity of the LISM. Since
the recombination rate of ionized  hydrogen is proportional to the product of
the  electron and  proton densities,  i.e.   effectively to  $N_{e}^2$ or  to
$N_{p}^2$, a given luminosity in Lyc photons can maintain ionized a volume of
plasma inversely proportional  to the square of the density.  To see how this
affects the range of  a Lyc photon, we can express this  relation in terms of
the  Str\"omgren  radius (the  radius  of  a  fully  ionized  plasma  sphere
maintained by a specific ionizing source). The Str\"omgren radius must vary as
$N_{e}^{-2/3}$ (or  $N_{p}^{-2/3}$). In an inhomogeneous  medium, this radius
can be  used as an upper  limit to the mean  free photon path,  if $N_{e}$ is
taken as the electron density of the low density phase of the ISM. An O3 star
yields a Str\"omgren radius of 140 pc in a homogeneous medium of mean density
$N_{p}$=1~cm$^{-3}$   (calculation   scaled   from   values   in   Vacca   et
al. 1996). The corresponding radius in a medium of $N_{p}$=0.1~cm$^{-3}$ will
be 650 pc, so  that a source equivalent to 4 O3 stars  will yield a radius of
$\sim$1~kpc in  a medium  with this density.   The net  result is that  in an
inhomogeneous  medium  an ionizing  source  with  a  strong flux  effectively
``burns" its  own long free path through  the low density phase;  this is the
same physical effect  as that analyzed by Miller \& Cox  (1993), though it is
certainly  applicable  to  more  circumstances  than  those  caused  only  by
supernovae (see Mckee \& Ostricker 1997) in yielding the initial low 
density ionized phase.

We are aware that there are almost certainly other sources of energy input
into the DIG, which were cited in Paper I. These have been inferred  from 
measurements of emission line ratios, notably [\sii]/\ha,[\nii]/\ha, [\sii]/[\nii]
 and [\oiii]/\ha, as discussed by a number of authors, including
Reynolds (1984), Reynolds, Haffner \& Tufte (1999), Greenawalt, Walterbos \& Hoopes 
(1997), Hoopes \& Walterbos (2000), Collins \& Rand (2000), and also 
from the presence of non--thermal velocity widths of emission 
lines (e.g. by Minter \& Balser 1999;  Tufte, Reynolds \& Haffner 1999).
In the present context, how\-ever, we are concentrating our attention on what
we believe to be the principal ionizing source of the DIG, i.e. Lyc 
escape from \hii\ regions, and our results tend to confirm that this approach 
is useful in practice.

In modelling  an external galaxy  to see whether  the OB stellar  sources can
account  for  the  observed  surface  brightness morphology  of  the  diffuse
\ha--emitting gas,  it is not  possible observationally to  resolve structure
within  the  ionized  gas  (nor  in  the neutral  gas).  This  means  that  a
microscopically realistic model of how the inhomogeneity affects the transfer
of ionizing radiation  cannot be tested fairly. In the  present paper we have
adopted a ma\-cros\-co\-pic approach, assuming large--scale homogeneity, so that an
absorption coefficient with an effective  mean value can be applied on scales
larger than tens of parsec. The range of trial values for this coefficient is
based on  simplified general transfer  calculations 
with realistic photon fluxes and gas densities. This method has enabled us to make
useful  comparisons between theory  and observation,  which go  as far  as is
warranted at  present by  the data,  notably in the  absence of  \hi\ surface
density maps at high spatial resolution.  The basic aim here is to go at least far
enough to  show whether or not the  hypothesis of the OB  associations as the
chief ionizing sources for the DIG can be discarded.
\section{The data used for the models of NGC~157}
In Paper~I we published a  careful analysis of the observed diffuse \ha\ from
the discs of 6 galaxies, but we  have chosen here only one of these, NGC~157,
to model the DIG surface brightness  distribution in \ha.  This is because it
is the only  one of our objects for  which we have an \hi\  map with anything
approaching fully adequate angular resolution.

\subsection{The \ha\ image}
The observations for this image, and their analysis, were described in
Rozas et al. (1996a). The galaxy was observed at the 4.2 m William Herschel
Telescope on La Palma, under good seeing conditions. The seeing in the reduced
final image was 0.8\arcsec. A luminosity function in \ha\ for the \hii\ regions
was extracted  and presented in Rozas et al. (1996a), after cataloguing 
708 \hii\ regions   in total.   The criterion for considering an image feature as 
an \hii\ region was that it must contain at least nine contiguous pixels 
each with an intensity of at least three times the r.m.s. noise of the 
local background (see  Rozas et al. 1996a for further details).

In Paper~I  we showed the surface brightness 
distribution of the diffuse \ha\ component, and inferred its integrated
luminosity, its radial distribution within the galaxy  and the fraction of the 
total galactic \ha\ in diffuse emission, as a function of radius and
in the galaxy as a whole. 

The method employed to separate the \ha\ flux emitted from the \hii\ regions from
that of the surroundings has benefitted from our previous cataloguing of the 
\hii\ regions. In the first place, we constructed masked images in which 
the \ha\ surface brightness 
of those pixels occupied by the catalogued \hii\ regions was set to zero (i.e. 
we assumed that there is no diffuse emission over each \hii\ region), considering
a circular approximation for the shape of each \hii\ region. After this, to take into
account that \hii\ regions are not truly spherical, a surface brightness cut--off 
equal to 73~pc~cm$^{-6}$ was employed over the masked images to avoid including 
in the DIG any residual emission from the \hii\ regions. A lower estimate of the
total \ha\ emitted from the DIG  of the galaxy was then obtained by integration of this 
masked image. The upper estimate was obtained in a similar way, but 
including an estimate for the diffuse emission above each \hii\ region. 
The detection thresold for the DIG of NGC~157 was 1.0 pc~cm$^{-6}$.
Detailed explanations of the isolation procedure and uncertainties in the measurements 
of the total \ha\ emission from the DIG are given in Paper~I.

As stated in paper~I, we have not made any correction, either for the
dust extinction within the emitting galaxy, or for Galactic extinction.

\subsection{The \hi\ map}
NGC~157 was observed by Ryder et al. (1998),  with the VLA
using the C and D configurations. The on--source exposure time with the 
C--array was 3.7 hours, and with the D--array 2.8 hours. The two data 
sets were processed separatedly and  combined. The total velocity range
covered was 581 km s\me\ with a channel increment of 5.2 km s\me. Two
 distinct data cubes were produced, 
using different weightings in the imaging process (see Ryder et al. 1998
for a detailed description of the observations, the 
reductions and the processing tasks). Here we used the \hi\ surface density 
map  obtained from the cube produced using uniform weighting, which yields 
better angular re\-so\-lu\-tion than natural weighting. The beam width at half power
for the map used here is 18\arcsec$\times$12\arcsec\ and the r.m.s. noise
1.3 mJy beam\me.

\section{Descriptions of the models for the propagation of the ionizing photons}
In the present section we explain the simplified theoretical models which
predict the ionizing photon distribution in the disc of the galaxy NGC~157,
assuming that they are produced within and escape from the catalogued \hii\
regions. Different laws governing the escaping photon fraction and the
absorption as they transit the ISM yield a set of models.
In these models, each measured \hii\ region is taken as an ionizing 
source and  fraction of the measured ionizing flux within each region
is taken as the luminosity of the escaping flux, available to maintain the 
ionization of the DIG. The \ha\ surface brightness of the DIG is com\-pu\-ted
by allowing the Lyc flux from a source to radiate within a thin disc in an
inverse square regime of dilution. In the simplest case, the fields from all
the observed \hii\ regions are co--added and the result compared to the observed
surface brightness distribution. Implicit in this procedure is that the
DIG surface brightness in \ha\ is linearly proportional to the underlying
Lyc field, which is the same as assuming that the disc consists of a slab 
of HI, whose properties do not vary from place to place.  It is clear that the 
observed \ha\ surface brightness must be the projection of a 
three--dimensional DIG column onto the plane of the galaxy observed (if it is
face--on). Given
that the scale--height of the DIG,  as shown within our Galaxy (Reynolds 1989) is of order 1 kpc, and that even the largest \hii\ regions do not project
more than some three  hundred pc  out of the mid-plane, the simplified geometry
employed here will yield a useful result for the integrated columns 
perpendicular to the plane viewed in a generally face--on direction. 

A second step in bringing the model into better relation with the
observations is to use the observed \hi\ surface density distribution to
modulate the Lyc flux, taking a uniform mass absorption coefficient to 
represent atte\-nua\-tion by gas plus dust extinction, and a third 
step is to take the product of the previously computed Lyc field and 
the \hi\ surface density for the predicted \ha\ distribution, assuming 
that the resulting surface brightness is the result of the irradiation 
of local neutral hydrogen by an incident ionizing field. The results 
from this type of modelling are
expected to be useful approximations provided that the ionized column density
is not a large fraction of the neutral column density, an assumption which we
tested for NGC~157, and which holds everywhere in the disc outside the \hii\ 
regions (see Sec~\ref{comparing}). 

An interesting invariance relation which helps to a\-ccount for the 
relative lack of apparent structure in the DIG, in spite of the assumed 
inhomogeneity of the clouds which compose it, and which also helps to account
for the fact that an essentially two--dimensional approximation can satisfy
a three dimensional flux distribution, is that the emission measure in \ha\ 
emitted by an optically thick cloud irradiated from outside is independent of
the cloud density, and depends only on the intensity of the irradiating field.
We can show this quite directly. In an optically thick case, where all the
incident photons are absorbed, the intensity of the recombination radiation
emitted (in the present case \ha)  can be related to the incident ioni\-zing 
radiation, using the ``on the spot" approximation (Osterbrock 1989), in which
any ionizing photons emitted by the ionized gas are reabsorbed locally. In 
this case, the number of ionizations per unit volume of the cloud in which the external 
flux has penetrated will be equal to the number of recombinations to excited
levels of the hydrogen:

\begin{equation}
N_{H^\circ}\int_{\nu_\circ}^{\infty}F_{\nu} \,a_\nu\,d\nu=N_{e}\, N_{p}\,\alpha_B(H^\circ,T)
\label{equil}
\end{equation}
where the atomic hydrogen density, the electron density and the proton density
are denoted respectively by $N_{H^\circ}$, $N_{e}$ and $N_{p}$, where
$\alpha_B(H^\circ,T)$  is
the recombination coefficient to excited levels of hydrogen, $F_{\nu}$ is the number 
of incident photons per unit area, time and frequency interval, and $a_\nu$ is 
the ionization cross--section for hydrogen by photons with energy $h\nu$ above the 
Lyman limit $h\nu_{\circ}$. If we integrate Eq.~\ref{equil} down to the depth 
within the cloud within which all the incident radiation is absorbed
(the optically thick case) using the equation of radiative 
transfer in the form:
\begin{equation}
dF_\nu/ds=-N_{H^\circ}a_\nu\,F_\nu 		
\end{equation}
we have:
\begin{equation}
\int_{\nu_\circ}^{\infty}\bigl\{\int_{F(s=0)}^
{F(s=L)}-dF_\nu\bigl\}\,d\nu=\alpha_B(H^\circ,T)\,\int_0^L N_{e}\, N_{p} ds 
\label{integral}
\end{equation}

The integrated intensity of the emitted \ha\ from a co\-lumn of 
length $L$ along which the electron and proton densities are given by  $N_{e}$ 
and $N_{p}$ respectively is proportional to the integral on the 
right hand side of Eq.~\ref{integral}, con\-ven\-tio\-nally termed the emission measure, 
EM, of the column. We can rewrite Eq.~\ref{integral}, incorporating 
the appropriate numerical constants, in the simpler 
form;
\begin{equation}
F(s=0)=A\,\, \alpha_B(H^\circ,T)\,\, {\rm EM\text{(in pc cm}^{-6}\text{)}}
\label{em}
\end{equation}
where $F(s=0)$  is the incident flux in the surface 
of the cloud,  
$A$ is a constant, whose numerical value for $F(s=0)$ in units of 
phot. s\me cm$^{-2}$  is $3.08\times10^{18}$
and $\alpha_B(H^\circ,T)$ does not show strong dependence on $T$ 
in the range normally found in the warm ionized ISM 
($\alpha_B\sim2.59\times10^{-13}$ cm$^3$~s\me). 	

Eq.~\ref{em} implies that irrespective of the total column depth
of the cloud, if it is optically thick to ionizing radiation it will develop 
an emitting skin for which the \ha\ intensity will be proportional to the
ionizing flux. Thus, seen from above, a set of \hi\ clouds with varying hydrogen
column density bathed from the side in uniform ionizing radiation will all 
show the  same surface emissivity. This result accounts in large part for 
the uniform appearance of the diffuse \ha\ from a galactic disc, in spite
of the intrinsic clumpiness of the substrate. This result does not
hold for an optically thin cloud, so that the net diffuse emission from a 
projected disc should not be entirely free from local variations. The result
above also implies that, if the ionization of the DIG is indeed due to OB stellar
clusters, the observed mean EM in the disc re\-la\-ti\-ve\-ly far from ionizing sources
should show a rough proportionality to the \hi\ column density. This is because 
if the \hi\ is in discrete clumps, optically thick to Lyc (but not in general to 
\ha), the integrated EM averaged over a column should be proportional to the 
integrated surface area of clumps per projected column cross--section, which 
will be a rising function of the integrated column density in HI, although the 
exact relation will depend on the details of the clump distribution. 
These considerations 
for the relation of the observed EM of the DIG to the gas distribution in the 
disc are valid for whichever detailed hypothesis we adopt about the relation of
the observed \ha\ in the \hii\ regions to their ionizing luminosities in Lyc. 

\subsection{Constant escape fraction (case A)}
\label{constant}
The simplest assumption to make when considering 
\hii\ regions as the zones of origin for the photons which io\-ni\-ze the DIG is
that a constant fraction of the Lyc from the OB stars in any \hii\ region 
escapes from the region into the surrounding diffuse gas. A number of previous
stu\-dies have suggested that a significant fraction of the total ionizing 
luminosity of OB stars can and does escape from their surrounding \hii\ regions, 
and is available to io\-ni\-ze the diffuse ISM. This conclusion is supported by
the geo\-me\-tri\-cal correlation observed between the sites of \hii\ regions and the 
surface brightness of the DIG (Ferguson et al. 1996). The escaping flux from a
given \hii\ region was not quantified by Ferguson et al. (1996)  who did not 
produce a photometric catalogue of \hii\ region luminosities, though 
their estimate of the total flux required to io\-ni\-ze the DIG was 
comparable to our estimates in Paper~I, i.e.  of order 50\% of the 
total for the whole disc. McKee \& Williams (1997), estimated that 
approximately 30\% of the ionizing photons 
observed within the \hii\ regions of the Galaxy, defined by the extent of 
their cm wavelength continuum emission, in fact escape from the \hii\ regions.
In general, the number of ionizing photons absorbed in an \hii\ region observed
in \ha\ will be higher than that absorbed only within the bright central
zones, which define the region at cm  wavelengths, but it is not apparent how
to define reliably a core--halo boundary in \ha. We chose a 30\% constant
escape fraction as our starting point for the ``constant escape fraction models"
of NGC~157. In these models, each measured \hii\ region is taken as an ionizing 
source and  a constant fraction of the measured ionizing flux within each region
is taken as the luminosity of the escaping flux available to maintain the 
ionization of the DIG. The \ha\ surface brightness of the DIG is computed
by allowing the Lyc flux from a source to radiate within a thin disc in an
inverse square regime of dilution. 
\subsection{Density bounding at high luminosity (case B)}
\label{limitdens}
In Paper~I we outlined a possible  model for the escape  of ionizing photons
from \hii\ regions, in which essentially all the Lyc escapes from the regions
of  highest   luminosity,  i.e.  those  with  observed   luminosity  in  \ha,
L$_{H\alpha}$,  greater  than a  critical  transition  value L$_{Str}$ (the 
``Str\"omgren transition luminosity'').  This
follows from a  series of studies in which  evidence has accumulated pointing
to the fact  that \hii\ regions of very high  luminosity are density bounded,
and  that  an  increasing  fraction  of  their Lyc  fluxes  escape  as  their
luminosities increase (Rozas et al.  1998; Beckman et al. 2000). 
We have developed a general parametric formulation showing the dependence of
the escaping ionizing  luminosity as a function of  the physical parameters of
the \hii\ regions with \ha\ luminosity greater than L$_{Str}$, whereas no
escape of Lyc is assumed for \hii\ regions in which  L$_{H\alpha}<$L$_{Str}$
in this case B. 

Let N$_{H\alpha}$  be the number  of ionizing photons  per unit time  from OB
associations which  io\-ni\-ze the surrounding  cloud and are  down--converted in
\ha\   emission  within   an  \hii\   region\footnote{The   relation  between
N$_{H\alpha}$  (number of ionizing photons per unit time 
from the OB associations down--converted in
\ha\   emission)  and  the observed  luminosity  in  \ha\  of an  \hii\  region,
L$_{H\alpha}$ is  N$_{H\alpha}= n {\rm  L}_{H\alpha}/(h\nu_{H\alpha})$, where
$h\nu_{H\alpha}$ is  the energy  of an \ha\  photon and n=2.233  Lyc photons/
H$\alpha$ photon  for T$\sim$10$^4$ K (Osterbrock 1989).}.   Assuming that an
\hii\ region contains gas in small condensations or clumps with electron density
$\rho_c$, N$_{H\alpha}$ can be expressed as: 
\begin{equation} 
{\rm N}_{H\alpha}=\frac{4\pi}{3} n \alpha^{eff}_{H\alpha} {\rho_c}^2 \phi \, {\rm
R}_{s}^3 
\label{n6} 
\end{equation} 
where $\alpha^{eff}_{H\alpha}$  is  the
effective \ha\  recombination coefficient, $\phi$  is the filling  factor (or
fraction of the total volume occupied by the condensations, as explained above),
$n$ is the number of Lyc photons necessary to yield one \ha\  photon
($n$=2.233  Lyc photons/H$\alpha$ photon  for T$\sim$10$^4$ K; Osterbrock 1989)
and ${\rm R}_{s}$ is the Str\"omgren radius of the
\hii\ region.

The io\-ni\-zed mass of the \hii\ region can be written as:
\begin{equation}
     {\rm M}_{H^+}=\int{m_{H} \phi \rho_c dV}=\frac{m_{H}}{n \alpha^{eff}_{H\alpha}} 
      \frac{{\rm N}_{H\alpha}}{\rho_c}\,
  \label{mas6}
\end{equation}
where $m_{H}$ is the mass of an atom of hydrogen and the integral is performed over the
entire volume ($V$) of the \hii\ region.

Density bounded \hii\ regions are fully io\-ni\-zed, so for these regions the mass of the
placental cloud (M$_{cl}$) is equal  to the mass of io\-ni\-zed hydrogen in the \hii\ region.
Combining  Eqs.~\ref{n6} and \ref{mas6} and normalizing the quantities (which will be noted
with $\hat{\,}\,\,$) by their value at the Str\"omgren transition, we find that
 
 \begin{equation}
     \hat{\rm N}_{H\alpha}=\frac{{\rm N}_{H\alpha}}{({\rm N}_{H\alpha})_{Str}}=
      \frac{\rho_c}{({\rho_c)}_{Str}} \frac{{\rm M}_{cl}}{({\rm M}_{cl})_{Str}}=\hat{\rho_c} \, \hat{\rm M}_{cl}\,
  \label{nha}
  \end{equation}
This equation holds for density bounded \hii\ regions.
 
In Beckman et al. (2000) a simplified scaling calculation is presented, 
in which the required relation between cloud--mass and the ionizing 
luminosity (L$_i$) emitted by a young cluster of stars which has formed 
in a placental cloud of mass M$_{cl}$ is given by
\begin{equation}
{\rm L}_i=k ({\rm M}_{\star})^{\alpha} = k j^\alpha ({\rm
 M}_{cl})^{\alpha\epsilon} \,,
\label{li}
\end{equation}
where  $k, \,j$  are constants  that  depend principally  on the  metallicity
(Beckman et  al. 2000) and  M$_{\star}$ is the  total ste\-llar mass in  the OB
association.  The  index $\alpha$ represents a  parametrized relation between
L$_{i}$  and the stellar  mass ($M_{\star}$)  in the  cloud, while  the index
$\epsilon$  is the  parametrized relation  between the  the stellar  mass and
M$_{cl}$.   The condition  that the  rate of  production of  ionizing photons
fully  io\-ni\-zes  and  overflows  the  placental  cloud  implies  that  $\alpha
\epsilon$  should  be  greater than  unity  (see  Beckman  et al.   2000  for
details). Observational evidence showing that $\alpha \epsilon$ should indeed
be greater  than unity  was reviewed in  Beckman et  al. (2000), and  a value
close to 1.6 was adopted as the best estimate.
Combining Eqs.~\ref{li} and \ref{nha} and normalizing to $({\rm L}_i)_{Str}$:
 \begin{equation}
    \hat{\rm L}_i=(\hat{\rm M}_{cl})^{\alpha\epsilon}=
    (\hat{\rm N}_{H\alpha})^{\alpha\epsilon} \hat{\rho_c}^{-\alpha\epsilon}
  \end{equation}
Under these conditions we can compute the 
escape fraction ($f$) of  ionizing photons down--converted in \ha\ (N$_{H\alpha}$)
which escapes from the most luminous regions as a function of the observed
luminosity in H$\alpha$:
\begin{equation}
\begin{array}{ll}
f=&\frac{\hat{\rm L}_{i}-\hat{\rm N}_{H\alpha}}{\hat{\rm N}_{H\alpha}}=
    (\hat{\rm N}_{H\alpha})^{\alpha\epsilon-1}
    \hat{\rho_c}^{-\alpha\epsilon}-1=\\
  & \\
  & =(\hat{\rm L}_{H\alpha})^{\alpha\epsilon-1} \hat{\rho_c}^{-\alpha\epsilon}-1\\
\end{array}
\label{fi}
\end{equation}
To estimate $f$ it is necessary  to know how the clump density varies 
with the \hii\ region luminosity. Measurements of \ha\ surface 
brightness alone will not allow us to detect whether $\rho_c$  is 
increasing as the luminosity increases, but do tell us (see Rozas et 
al. 1996b, 1998) that the product $\rho_c \phi^{1/2}$, which is equal to 
$<{\rm N}_e>_{rms}$,  increases by a factor close to 2 
between the Str\"omgren luminosity and the luminosities of the 
brightest regions, and enables us to use 
Eq.~\ref{fi} to set bounds to the predicted escaping photon flux from 
density bounded regions, as a function of luminosity. An example of this is 
\begin{figure}
\begin{center}
\resizebox{8cm}{!}{\includegraphics{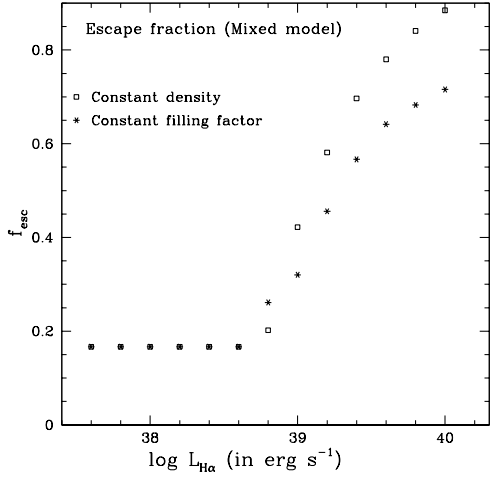}}
\parbox[e]{9cm}{\vspace{-1cm}\caption{Fraction of the total 
ionizing photon flux within an \hii\ region, which escapes from
it in the mixed model C (see text for details). The relation of 
f$_{\rm esc}$ (in the figure) to $f$ (Eq.~\ref{fi}) is given by  
f$_{\rm esc}$=$f$/(1+$f$). Approximately 80\% of the total ionizing radiation  
escapes from the most luminous \hii\ regions in NGC~157 in this 
model (i.e. nearly 4 times the ionizing luminosity down--converted 
to \ha\ emission within the region).
}\label{fiescp}}
\end{center}
\end{figure}
given in Fig.~\ref{fiescp} (for luminosities greater than L$_{Str}$)
 in which the two limiting cases of varying
$\rho_c$  and fixed $\phi$ or varying $\phi$ and fixed $\rho_c$ are 
taken and used to predict the escape fractions of ionizing photons 
versus luminosity. Fig.~\ref{fiescp} shows f$_{\rm esc}$, which is the 
fraction of the total ionizing photon flux which escapes from the 
\hii\ region. Note that if $f$ (Eq.~\ref{fi}) is the fraction of ionizing photons 
down--converted to  \ha\ (N$_{H\alpha}$) which escapes from the \hii\ region, the 
relation between the two quantities is given by f$_{\rm esc}$=$f$/(1+$f$).

In the models, to maintain the family of calculated cases as simple as possible,
we maintained $\alpha \epsilon$ at a constant value of 1.6, and  $\rho_c$ constant,
varying $\phi$ in order to reproduce the range of observed surface brightness
in \ha\ v. L$_{H\alpha}$ found for the regions in NGC~157 by Rozas et al.~(1996b). 
We note that at the highest luminosities the escape fraction rises to values
well above 0.5, which means that more than 50\% of the Lyc photons escape from
the region. We can use the interpolated curve in Fig.~\ref{fiescp} to compute 
an escape fraction from each \hii\ region in the catalogue of regions for 
the galaxy, so that the absolute luminosity in ionizing photons escaping 
from each region is then tabulated, and can be used as input to model 
the production of the diffuse \ha\ in the same way as for the models 
with constant escape fraction, as explained in the preceding Sec.~\ref{constant}.

\subsection{Mixed models (case C)}
\label{mixtos}
In this model we assume that the major contribution  of ionizing photons 
available to io\-ni\-ze the DIG comes from the most luminous \hii\ regions 
(those with log~L$_{H\alpha}\gtrsim$38.6, in erg s\me), whose escape fraction
is given by Eq.~\ref{fi}, but a minor contribution escapes from the less
luminous regions (with log~L$_{H\alpha}\,<$ 38.6, in erg s\me), for which
a constant escape fraction of ionizing photons is assumed ($f$=20\%).

Inspection of the \ha\ image of  NGC~157 showed us that the diffuse component
is  relatively strong  near  an  \hii\ region,  and  that this  observational
correlation appears to hold not only  for the most luminous regions, but also
for regions with  luminosities below L$_{H\alpha}$=L$_{Str}$. The implication
is  that  even those  regions  with lower  luminosities  are  leaky. Such  an
inference  is   entirely  plausible  given   the  clumpy  structure   of  any
interstellar cloud and,  in par\-ti\-cu\-lar, of the clouds which  make up the \hii\
regions. Proof of the clumpy nature  of the regions is found by comparing the
r.m.s. electron density in a region (Rozas et al. 1996b, 1999, 2000) obtained
from  a diametral  emission  measure in  \ha\  with in  situ measurements  of
electron density obtained via line  ratios (Zaritsky et al. 1994), as explained above in 
Sec.~\ref{intro}. 
Relatively crude models using  this structure give as a result  an 
escaping fraction for
the Lyc produced by their OB populations, a fraction which should vary li\-ttle
if   the   density   structure   of   a  region   depends   little   on   its
luminosity. Measurements of filling factors  show that, at any rate for
L$_{H\alpha}<$~L$_{Str}$ this condition appears  to hold rather well. However,
as  explained above  in Sec.~\ref{constant}  at the  highest  luminosities the
r.m.s. electron density shows higher  values (Rozas et al. 1996b, 1999, 2000)
which we interpret as due to  an increased effective filling factor, which in
turn  is  explained if  the  clumps near  the  stellar  sources become  fully
io\-ni\-zed,  releasing  more  Lyc  to  io\-ni\-ze  clumps further  out within  the
region. In this model each  in\-di\-vi\-dual clump behaves as an ionization bounded
or  a density bounded  system, according  to the  ionizing luminosity  of the
region, and the distance of the clump from the star cluster. The \hii\ region
as a whole then  starts to leak a higher fraction of  its Lyc luminosity when
the inner dense  clumps become density bounded, which  will occur for regions
of high stellar luminosity. Without  going into further detail, we have taken
this into  account empirically  in our termed ``mixed  models'' by using  a constant
escape fraction for L$_{H\alpha}<$~L$_{Str}$ and an increasing escape fraction
for L$_{H\alpha}>$~L$_{Str}$,  following Fig.~\ref{fiescp}.

\section{Comparing observations with the model predictions}
\label{comparing}
\subsection{Models with no computed in--plane extinction}
\begin{figure}[!]
\begin{center} 
\vspace{-4cm}
 \resizebox{9cm}{!}{\includegraphics{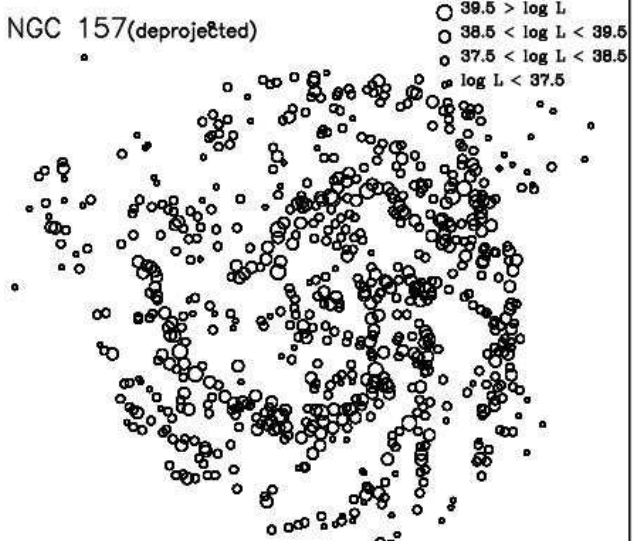}}\vspace{-1cm}
 \hfill  \parbox[b]{9cm}{  \caption{Representation of  the  positions of  the
 catalogued \hii\  regions in the  deprojected disc of NGC~157.  The vertical
 direction  of the  plot is  coincident with  the major  axis of  the galaxy.
 Symbols show  ranges of log  L$_{H\alpha}$.}  
\label{157repre}} 
\end{center}
\end{figure} 
The basis  for all the models  is the positional and  lu\-mi\-no\-si\-ty catalogue of
 all the  detected \hii\ regions  in NGC~157\footnote{For  the criteria of  the lower
 limit of  detectability the reader  is referred to  Rozas et al.  1996a.}. In
 Fig.~\ref{157repre} we  show this in  display form, i.e.  the  positions and
 luminosities of all the regions used in the models are indicated in this map
 deprojected into the plane of the sky. To use the regions as sources for the
 diffuse  component, we  first apply  the  escape factor  appropriate to  the
 model: constant   factor    (case   A),   pure    density   bounding   above
 L$_{H\alpha}$=L$_{Str}$ (case B), or mixed model (case C), to each and every
 one  of  the regions.  Treating  each  region as  a  point  source with  
 effective  Lyc luminosity  equal  to  that  corresponding  to  its  
 measured  \ha\  output,
 multiplied by the escape factor, we  then model the diffuse emission using a
 set of assumptions of increasing complexity.

\begin{figure*}
\vspace{-1.2cm}
\begin{center}
\resizebox{14cm}{!}{\includegraphics{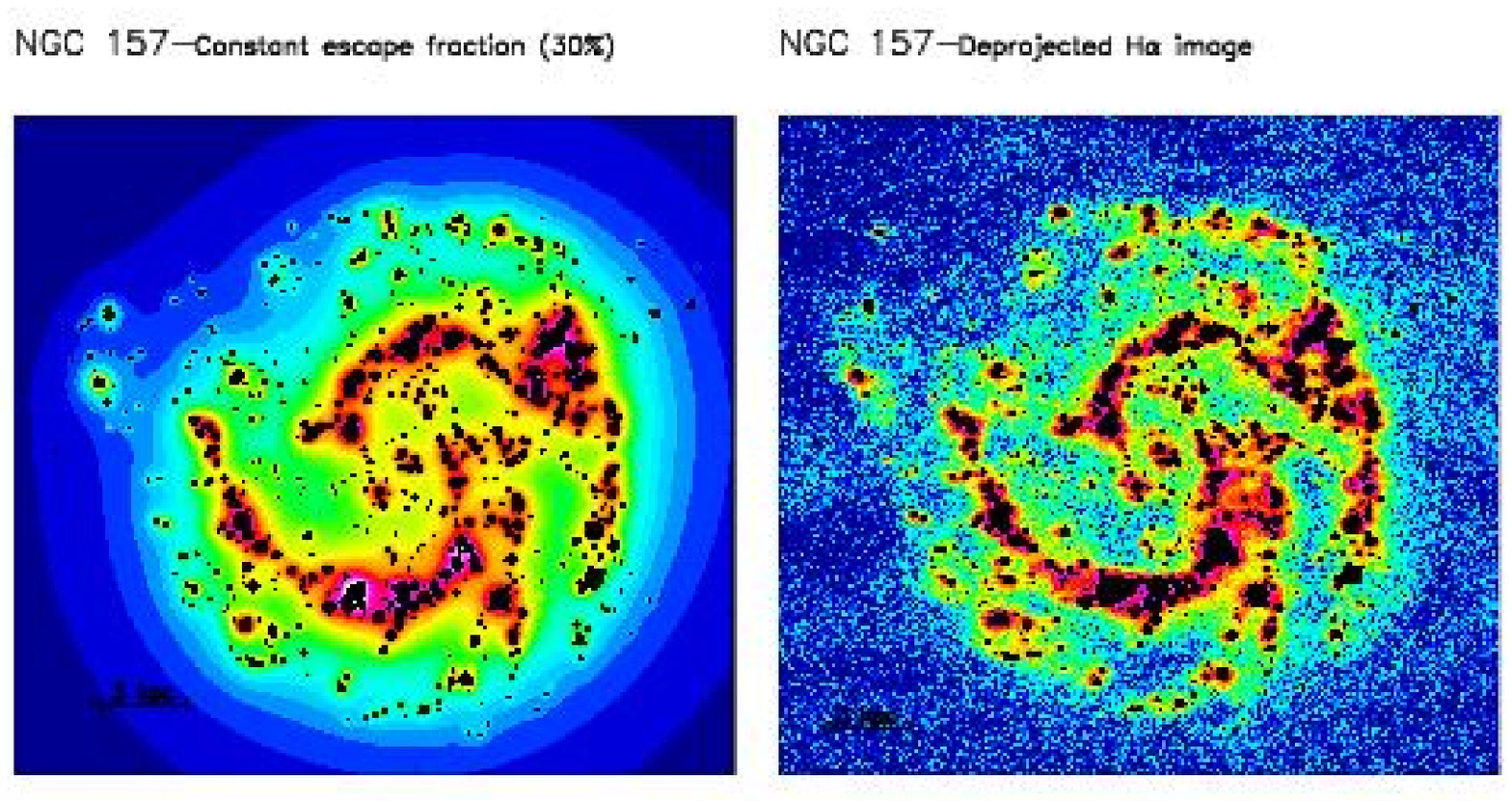}}
\parbox[e]{18cm}{\vspace{-2.5cm}\caption{ {\bf  Left:} Surface  photon density
in the  disc of  NGC~157 assuming  that a constant  fraction of  the ionizing
photons escape from each catalogued \hii\ region. \hii\ regions appear in the
figure  as black  circles. {\bf  Right:} Deprojected  \ha\ image  of NGC~157.
\hii\ regions  have been masked.  The lower \ha\ surface  brightness cut--off
applied  to this galaxy  in Paper~I for  measuring the  emission of  the DIG
(73$\times \cos{i}$  pc cm$^{-6}$, $i$=50\deg, de Vaucouleurs et al. 1991)  
has also been  applied here,  taking into
account  the  deprojection.  The  two  images  ({\bf left} and {\bf right})  have  been
normalized and plotted using the same colour scale.}
\label{modepro}}
\resizebox{12cm}{!}{\vspace{0.0cm}\includegraphics{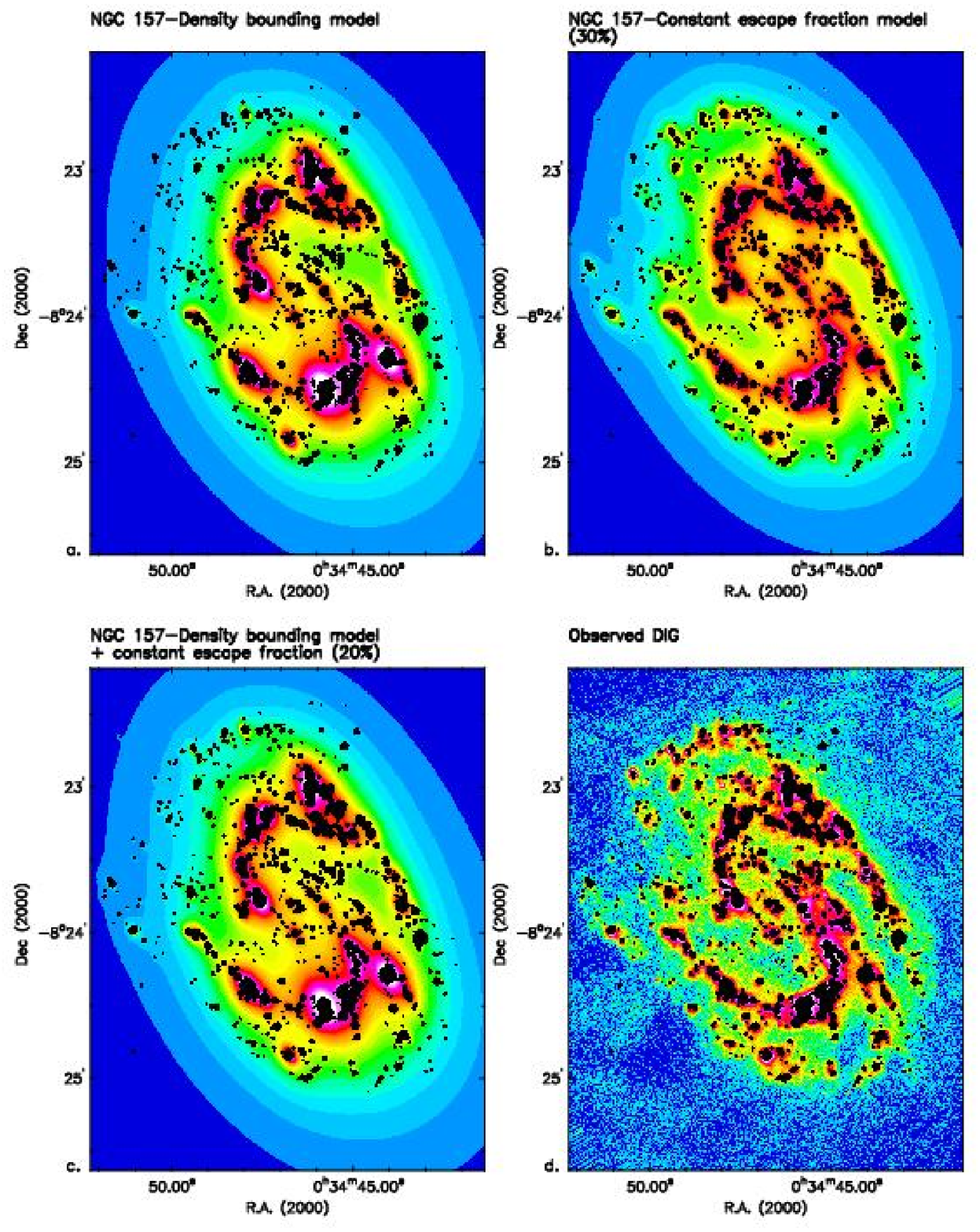}}
 \parbox[e]{18cm}{\vspace{-0.5cm}\caption{Colour    representation   of   the
 modelled ionizing photon density in  the disc of NGC~157, assuming that this
 distribution  is produced  by the  escaping photons  from the  most luminous
 regions alone  (case B,  {\bf a.}), from  all the catalogued  \hii\ regions
 (case  A, {\bf  b.}) or  from all  \hii\ regions  but with  different escape
 fractions (case  C, {\bf c.}). The  images have been normalized  to the mean
 value in the ellipse determined by  the observed \ha\ emission of the galaxy
 (see   Paper   I).   The  observed   DIG   is   shown   in  {\bf   d.}   for
 comparison (also normalized).}\label{mossinabs}}
\end{center}
 \end{figure*}

In this section we assume a uniform HI
layer of constant thickness.
In the simplest case, we compute a flux density $I_j(x,y)$  at a point
($x$, $y$)  in the disc due to a region ($j$) emitting $D_j$ escaping 
photons per second, via:
\begin{equation}
    I_j(x,y)=\frac{D_j}{4 \pi [(x-x_j)^2+(y-y_j)^2]}
\label{dilucion}
\end{equation}
where ($x_j$,$y_j$) are the coordinates of the region selected. We then sum over 
all the regions to obtain the final ioni\-zing photon flux density map from the
contribution of all \hii\ regions:
\begin{equation}
    I(x,y)=\sum_{j=0}^{N}I_j(x,y)\,
\end{equation}	
where $N$  is the  total number of  catalogued regions (weaker  regions would
make a negligible impact on the  result using any of the assumptions taken in
this work (and can genuinely be  neglected). A sample result of computing $I(x,y)$
in this way  is shown in Fig.~\ref{modepro}a where we  have taken the simplest
case, i.e.   a hypothetically constant  fraction of escaping photons  for the
$D_j$   values.    We   have   compared   this   computation   directly,   in
Fig.~\ref{modepro}(right) with the deprojection into  the plane of the sky of
the diffuse \ha\ measured  in NGC~157, and we can see that the  result is to say
the least encouraging. All the  broad intensity and morphological features of
the diffuse \ha\ are reproduced, although inspecting the fine details we 
can pick out
certain discrepancies. We should state now that in the rest of the discussion
rather than deproject the observed  diffuse light distribution into the plane
of the sky  we will project the model  into the plane of the  galaxy disc, as
this is an o\-pe\-ra\-tion  which of course is not affected by  low signal to noise
ratio.

\begin{figure}
\vspace{-2.2cm}
\begin{center}
\resizebox{9cm}{!}{\includegraphics{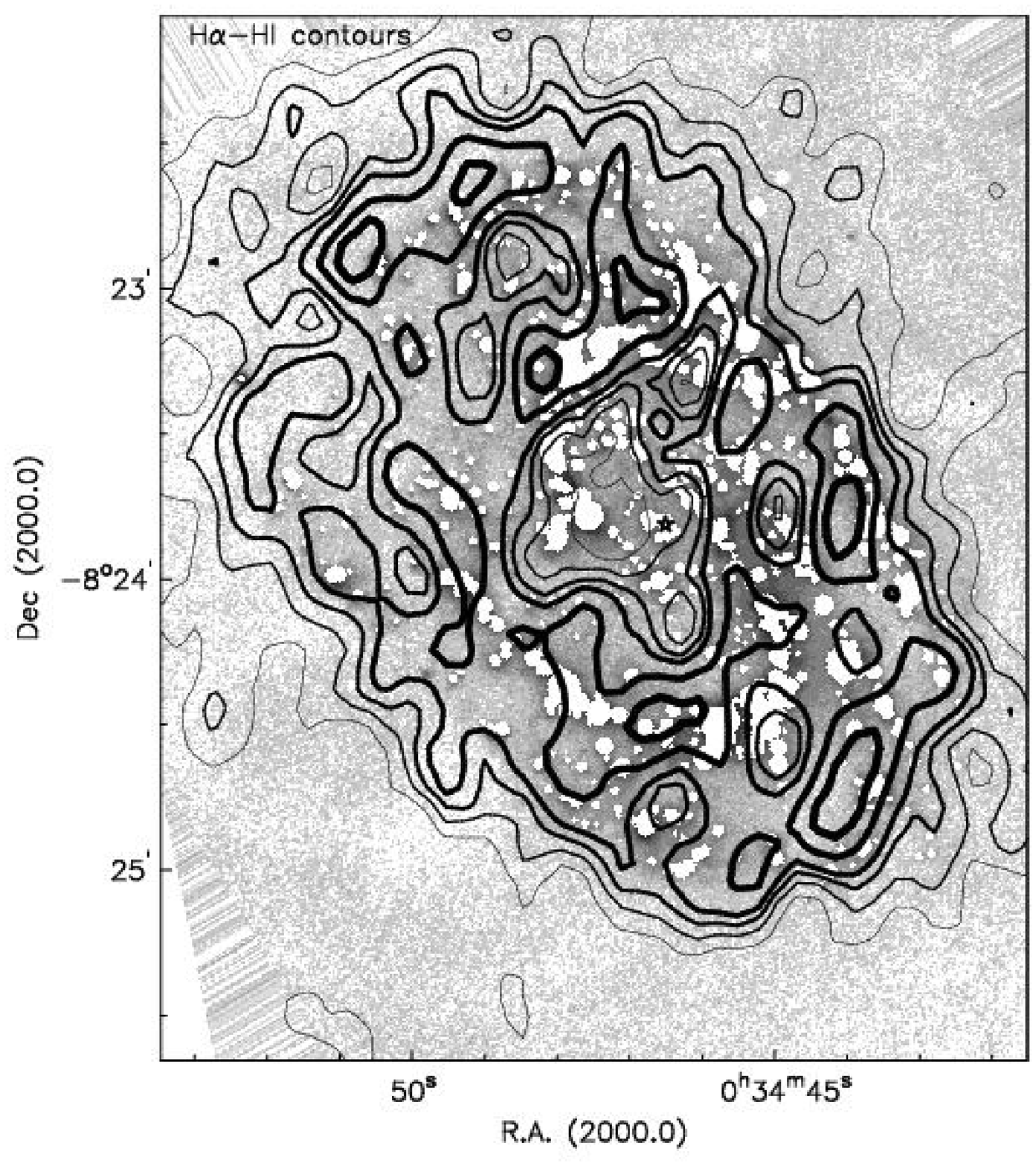}}
\parbox[e]{9cm}{\vspace{-0.3cm}\caption{\ha\  surface brightness  density of
the  observed DIG  in NGC~157  with  \hi\ column density  contours  superposed. 
The density contours are 1.45, 2.90, 4.35, 5.80, 8.70, 11.60 M$_{\odot}$/pc$^{2}$
(column density increasing from finer to thicker contours).
The two quantities (\ha\  surface brightness and \hi\ density) should  
be proportional in  the ``Sciama neutrino'' model.  We can
see that  (in spite of the difference in  resolution between the two  images) 
the \hi\ distribution does not correlate at all with the \ha\ emission 
by the DIG. Darker areas in the
image correspond to higher levels of  \ha\ emission by the DIG of the galaxy. 
Thicker contours represent the higher \hi\ column densities. The 
parallel shading in  the corners
of the image is an  artifact generated in the rotation of the image to obtain the 
co\-rrect orientation (the same effect is seen in  Figs.~\ref{mossinabs}d,~\ref{mosxh1}d, ~\ref{colden} and ~\ref{mosabs}).               
}\label{ha-h1}}
\end{center} 
\end{figure}

The fact that  the morphological agreement between model  and observations is
so good is  perhaps a little surprising, since the  conditions imposed are so
simple. Without  further refinement we can  use this result  to criticize the
hypothesis that the diffuse \ha\ is originated by decaying massive neutrinos,
although  this model  by Sciama  (1990) has  been already  found  wanting via
direct EUV observations (Bowyer et al. 1999). We can do this by comparing the
measured  distribution of \hi\  column density  in Fig.~\ref{ha-h1}  with the
\ha\ distribution and model prediction.  
The \hi\ density contours for NGC~157  (Ryder et al. 1998) have been overlaid
on the  \ha\ image  of the galaxy.   The angular  resolution is poor  by \ha\
standards, but  we can see clearly  that zones of high  surface brightness in
diffuse \ha\ do not  coincide with zones of high \hi\ column  density, a result which
contradicts the  predictions of the  massive neutrino decay  hypothesis.  
This lack of correlation is maintained if we smooth the \ha\ image to match the resolution of the \hi\
image. 
The leaky \hii\ region model, even in its simplest form as presented here, offers
a much better solution.

In  Fig.~\ref{mossinabs}  we  present  the  results of  applying  the  simple
geometrical  propagation  model to  leaky  \hii\  regions  in three  regimes:
constant  escape fraction, density  bounding only  for $L_{H\alpha}>L_{Str}$,
and  the  ``mixed model"  as  des\-cri\-bed  above,  compared with  the  observed
distribution of  the diffuse \ha.   It is clear  that the model  with density
bounding only  for $L_{H\alpha}>L_{Str}$  yields a worse  fit to  the surface
brightness than  the other two.  On the  basis of this result  we can already
begin to question the hypothesis put  forward in Beckman et al.  (2000), that
only those regions with $L_{H\alpha}>L_{Str}$ are density bounded.  This is
particularly clear for  the diffuse emission from the  group of \hii\ regions
forming the  uppermost arm  of NGC~157. Although  most of these  regions have
$L_{H\alpha}<L_{Str}$ the arm is su\-rrounded by diffuse emission, which is not
well reproduced  in the  ``density bounding" model  where Lyc  photons escape
only  from the  regions with  $L_{H\alpha}>L_{Str}$.  In  all the  models the
distribution of the diffuse emission is less confined to the zones around the
arms than in the  galaxy as observed, and this leads us  to conclude that the
underlying assumptions are rather too simplified.

\subsection{Models incorporating the observed \hi\ column density distribution}

\begin{figure*}
\vspace{-1cm}
\begin{center}
\resizebox{13cm}{!}{\includegraphics{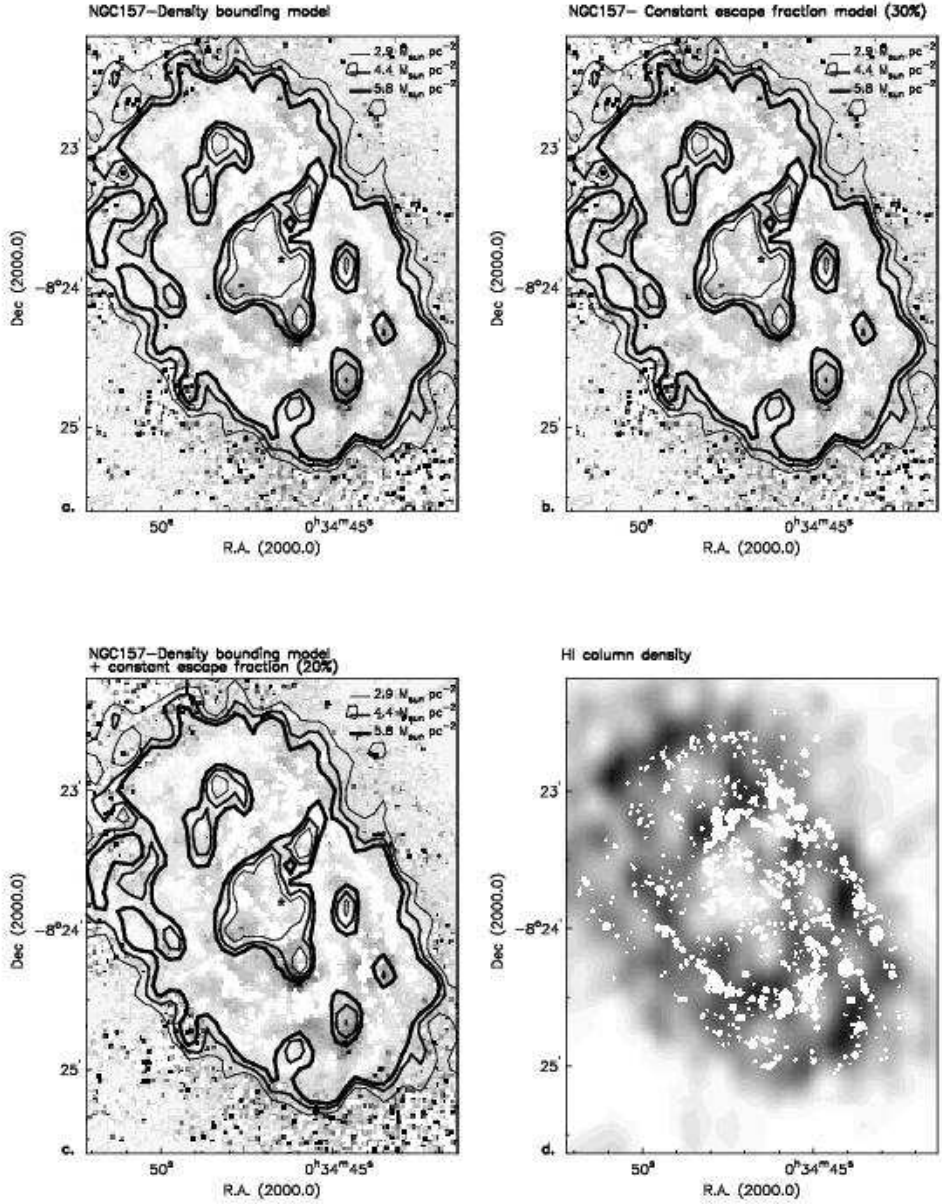}}
\parbox[e]{15cm}{\caption{First  three panels: ratio  of the  modelled photon
distribution, for the three  cases shown in Fig.~\ref{mossinabs}a,b,c, to the
observed DIG emission in \ha. Pixels have been binned to a new pixel size equal to 4.
Only the lowest \hi\ density contours have been superposed (cf. Fig.~\ref{ha-h1} and 
panel d.) 
There is good spatial coincidence of the zones where the models with an assumed
layer of \hi\ of constant thickness overestimate the emission (high ratios 
correspond to darker areas), with the ''holes" in the measured \hi\ distribution.
This distribution is shown in grey scale in the fourth panel, in which the darkest areas
correspond to the highest \hi\ column densities.}\label{mosraztesis}} 
\end{center} 
\end{figure*}

\begin{figure*}
\vspace{-1cm}
\begin{center}
\resizebox{12cm}{!}{\includegraphics{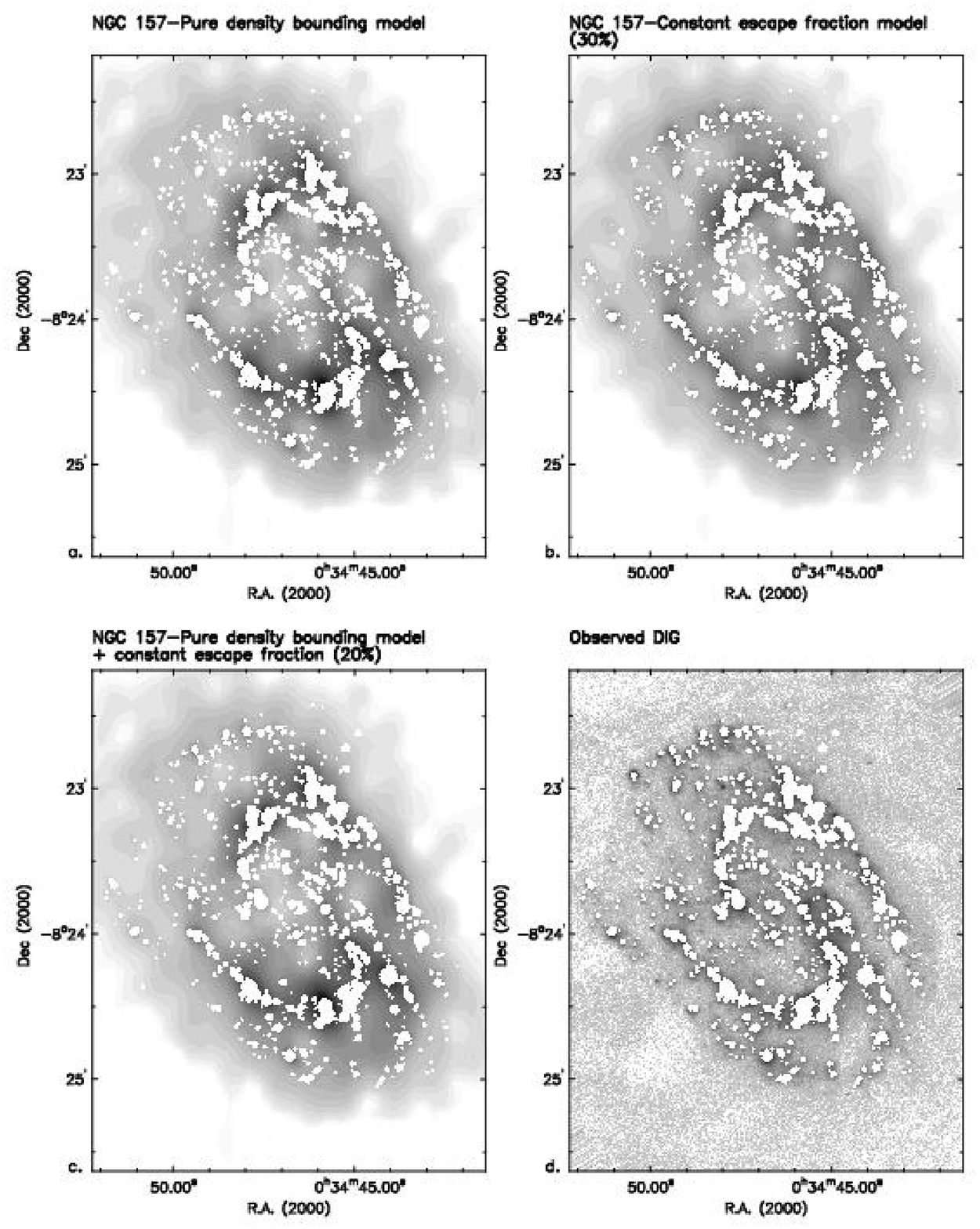}}   
\parbox[e]{15cm}{\vspace{-0.5cm}   \caption{Monochrome  representation  of   the
modelled diffuse \ha\ emission for the galaxy NGC~157. Each figure represents
the product of the photon distribution  and the \hi\ density at each point of
the  galaxy for  the cases  B, A  and C  shown in  Fig.~\ref{mosxh1}.  
All four figures have been normalized to the mean value in the ellipse determined by
the observed \ha\ emission of the galaxy and plotted using the same grey scale.
The
different resolutions  between the \ha\  and the \hi\  data does not allow  us to
extract a definitive  conclusion from these maps, but  the broad agreement in
the overall DIG distributions predicted with that observed, strongly favours
the hypothesis that photons leaking from  \hii\ regions do power the DIG (see
Sec.~\ref{discusion} for a discussion).  \label{mosxh1}}} 
\end{center}
\end{figure*}

One of the simplifying assumptions implicit in the results of Fig.~\ref{mossinabs}
is that the \hi\ which down--converts the escaping Lyc photons to \ha\ is in 
a disc of uniform thickness. We carried out a check to see whether at least 
part of the discrepancy between models and observations could be attributed
to this. In Fig.~\ref{mosraztesis}a,b,c, we show ratios of the modelled \ha\ surface 
brightness to the observed surface brightness, for the three cases presented in 
Fig.~\ref{mossinabs}. The ratios are those of Figs.~\ref{mossinabs}a,b
and c, respectively,
divided by Fig.~\ref{mossinabs}d, and smoothed to four times the original pixel
size. In Fig.~\ref{mosraztesis}d we have
shown the measured \hi\ surface density for comparison. We note that the zones
where the  ratio of predicted  to observed  \ha\ is high  in
Figs.~\ref{mosraztesis}a,b,c,
coincide with  the zones  where the  observed \hi\ column  density is  low in
Fig.~\ref{mosraztesis}d, which  implies that our  neglect of the  \hi\ column
density variations in  our model predictions of Fig.~\ref{mosraztesis}
   needs to be corrected.
We have performed  such a correction, again in the  simplest way possible, by
multiplying directly  the predicted Lyc field  in a model by  the \hi\ column
density.  The results are shown in Fig.~\ref{mosxh1} where we can see 
that, globally, the distribution of \ha\ from the DIG is better reproduced in 
general than in
Fig.~\ref{mossinabs}, in  that the diffuse  component is more intense  closer to
the  arms, and  less so,  further  away.  However  in one  aspect the  ``\hi\
corrected" models reproduce the oberved DIG less well than the previous ones,
i.e.   in the  zone around  the  centre of  the galaxy,  where the  predicted
intensity is  too low in  the \hi\ corrected  models. One reason for  this is
that  we have  simply assumed  that the  \ha\ emission  ought to  be linearly
proportional to the  \hi\ column density. This would  be a fair approximation
only if the \hii\  column density were small compared to that  of \hi; if not
\begin{figure*}
\begin{center}
\hspace{-0.9cm} \resizebox{16cm}{!}{\includegraphics{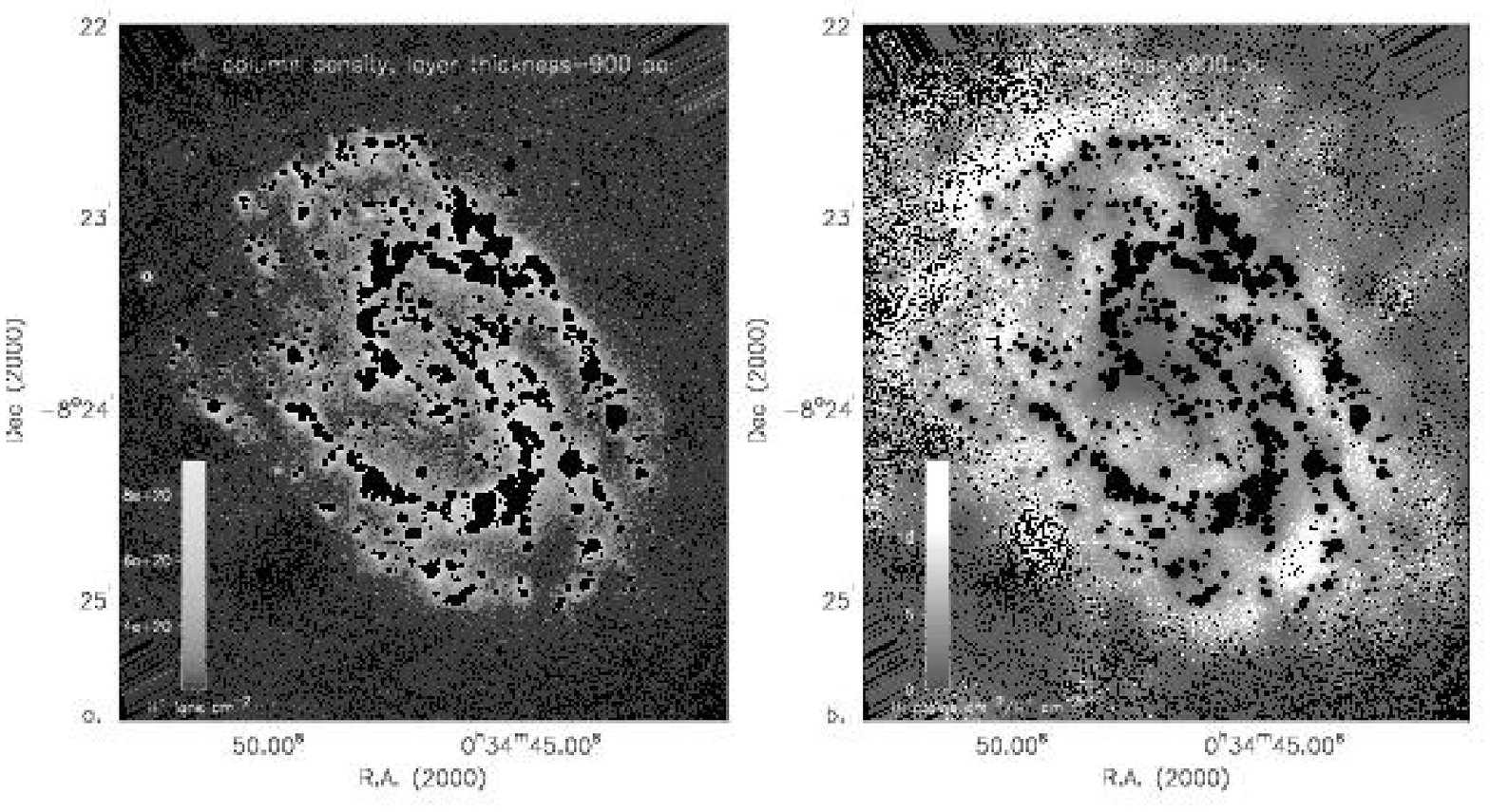}} \hfill
\parbox[e]{15cm}{\vspace{-0.2cm}\caption{{\bf  a.}  H$^+$  column density  map
assuming an effective thickness of the \ha\ emitting layer ($h$) equal to 900
pc.   Since   the  io\-ni\-zed   mass  per  unit   area  is  given   by:  $m({\rm
H}^+)=<N_e>_{rms}m_H h$  and the  emission measure EM$=<N_e>_{rms}^2  h$, the
relation between  the number  of io\-ni\-zed atoms  of H$^+$ per  cm$^2$, $n({\rm
H}^+)$, the EM  and h is given by:  $n({\rm H}^+)= \sqrt{\mbox{EM}} h^{1/2}$.
The mean value for the DIG is $\sim 5\times10^{20}$ H$^+$ ions cm$^{-2}$. For
other  scale heights  the  monochrome  bar calibration  changes  by a  factor
$(\frac{h}{900 pc})^{1/2}$. {\bf b.} Ratio of the \hi\ column density to the
H$^+$ column density  assuming an H$^{+}$ effective layer  thickness equal to
900 pc in the disc of  NGC~157.  In spite of the different image resolutions,
in the  DIG area  most distant from  \hii\ regions (where  surface brightness
gradients are nearly constant, see Paper~I) the figure will give a reasonable
estimate of the  ratio of the two column densities. In  these areas, the \hi\
column  density is  between 5  and  12 times  greater than  the H$^+$  column
density.}\label{colden}}
\end{center} 
\end{figure*} 
it would seriously under--estimate the  predicted \ha.  We carried out a test
for this, illustrated  in Fig.~\ref{colden}, in which we  first estimated the
\hii\  column density,  and then  compared  it with  that of  \hi.  Since  we
measure directly a surface brightness, i.e.  an emission measure (in \ha) our
estimate  of  the  column  density   cannot  be  unique,  without  a  further
assumption, which is the effective  thickness (h) of the \ha\ emitting layer.
Unfortunately the  result depends on the  square root of  this thickness, which
cannot  be measured.   In  Fig.~\ref{colden} we  have  used 900  pc for  this
thickness,   taking   the  ``Galactic   Reynolds  layer''   as  our   guide.
Fig.~\ref{colden}a maps the resulting \hii\ column density in the DIG, and in
Fig.~\ref{colden}b we have divided this into the observed \hi\ column density,
yielding a map of \hi/\hii. We see  from this map that the ratio is between 5
and 15 over most of the galaxy disc, but that near the centre it drops to low
values, so  that we would expect  our \hi--corrected models to  fall short of
the  observed  surface  brightness in  \ha\  near  the  centre, as  found  in
Fig.~\ref{mosxh1}.

One problem  in allowing for the \hi\  in this modelling process  is that the
angular resolution  in the \hi\ map  of NGC~157 is no  better than 12\arcsec,
which is a factor  of 15 lower than we would wish,  in order to make adequate
use of our  \ha\ map. In spite of  this we have gone one step  further in our
modelling, taking  into account the  absorption by the intervening  medium of
the propagating Lyc field. This is done by substituting Eq.~\ref{dilucion} by
\begin{equation}
\begin{array}{ll}
 I_j(x,y)&=\frac{D_j}{4 \pi [(x-x_j)^2+(y-y_j)^2]} \exp{(-\tau_{j})}=\\
           &  \\
           &=F_j \exp{(-\tau_{j})}
\end{array}
\label{ext}
\end{equation}
in which $F_j$ is the flux incident on unit path length at $(x,y)$ from a source
$j$, and $\tau_j$  is the optical depth along the path to $(x,y)$,  which is 
the integrated product of the effective volume absorption coefficient ($k$) and 
the volume density ($\rho_{HI}$)  along the path between the emitting 
\hii\ region $j$ and the point $(x,y)$. In implementing this model, 
the effective absorption coefficient is taken as an empirical parameter, 
which includes absorption by the \hi\ (the cause of the \ha\ emission) and 
also dust extinction, with an assumed  constant dust to gas ratio. 

Using 
Eq.~\ref{ext} we can express the
decrement, $dI_j/ds$ of ionizing radiation per unit length {\em ds} 
at the point $(x,y)$  by
\begin{equation}
\begin{array}{ll}
\frac{dI_j}{ds}&=\frac{dF_j}{ds}\exp{(-\tau_j)}+F_j \frac{d\exp{(-\tau_j)}}{ds}=\\
&\\
&=\frac{dF_j}{ds}\exp{(-\tau_j)}+(-k \rho_s)F_j
\end{array}
\label{absorcion}
\end{equation}
The first term here represents the decrement due to geometrical dilution, and
the second term the component due to absorption, both terms being intuitively
clear. In order  to get a reasonable value  of $k$ without three--dimensional
modelling of  the clumpy medium, we  take an empirical approach  based on the
observation that the diffuse component  extends to distances of order 0.5~kpc
from the concentrations of \hii\ regions.  With this starting point we take a
trial value of  $k$ as that which would  reduce a flux to 1\%  of its initial
value over a  range of 500~pc. Taking an a\-ve\-rage \hi\  scale height of 100~pc
as a constant  value over the disc and  a mean column density in  the disc as
that  observed: $8.6\times10^{20}$~cm$^{-2}$, we  find  a value  for $k$  of
5.3$\times10^{-22}$  cm$^{2}$,  or  more  generally  5.3$\times10^{-22}\times
z_{100}$ cm$^{2}$, where $z_{100}$ is the  value of the scale height in units
of 100~pc. In fact, this value for $k$ ought to be an upper limit, because the
condition that the flux be reduced by  a factor 100 over a distance of 500~pc
is  a stronger  condition than  that  required by  the empirical  map of  the
DIG.  In  our  numerical  models  we  took  values  of  $k$  ranging  between
$5.3\times10^{-24}$ and $5.3\times10^{-22}\times z_{100}$~cm$^{2}$.

\begin{figure*}
\vspace{-1cm}
\begin{center}
\resizebox{12cm}{!}{\includegraphics{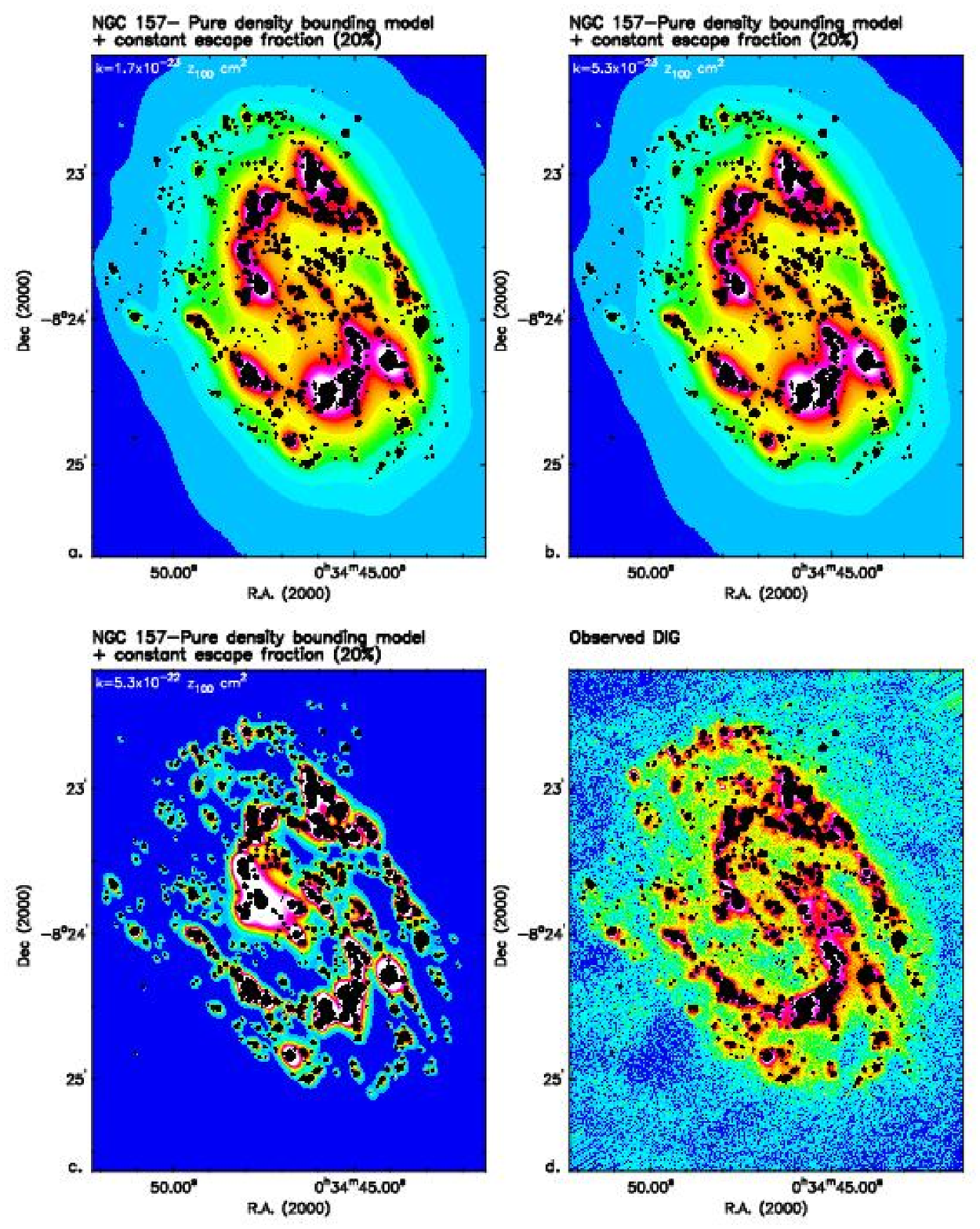}} 
\parbox[e]{17cm}{\vspace{-0.5cm}\caption{Same as Fig.~\ref{mossinabs} but for
models including  absorption by neutral  hydrogen. In this case,  all figures
assume case C (mixed model) for  the escape fraction of ionizing photons. The
effective volume absorption  constant for each case is shown  in the left top
corner of each representation.\label{mosabs}}} 
\resizebox{12cm}{!}{\vspace{-50cm}\includegraphics{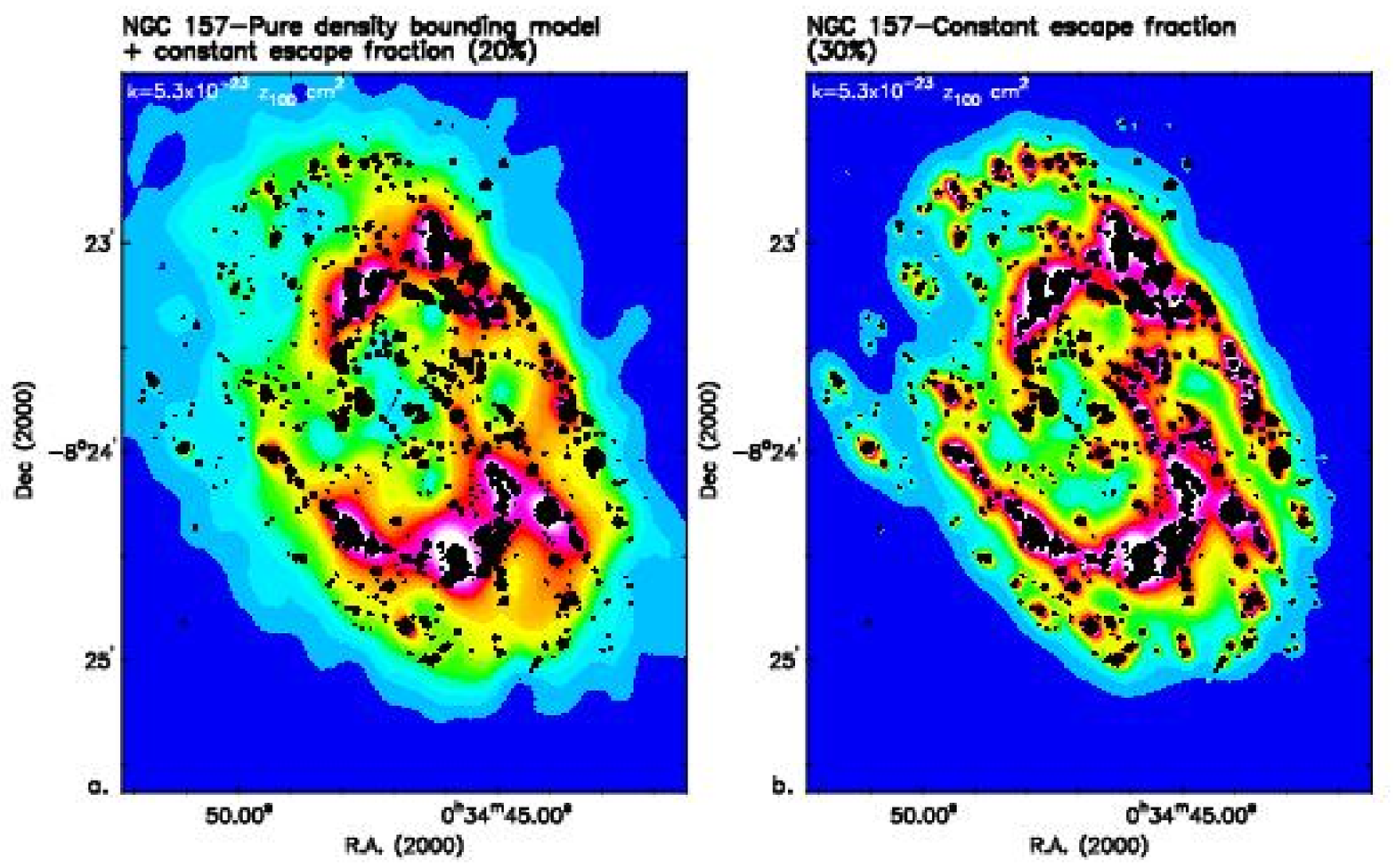}} 
\parbox[e]{17cm}{\vspace{-0.5cm}\caption{Same as Fig.~\ref{mosxh1}, but for 
models including absorption {\bf a.} for the mixed model and {\bf b.} for 
the constant escape fraction model.}\label{mosvarios}}
\end{center} 
\end{figure*}

Our models are of two kinds: mixed models in which the absorption coefficient
$k$ is allowed to vary, but in which  the \hi\ is assumed to take the form of
a disc of constant thickness, and  models in which an optimized value for $k$
is  combined with  the observed  \hi\ column  density map.  The two  types of
models are illustrated in Figs.~\ref{mosabs} and \ref{mosvarios}
respectively. 
In Fig.~\ref{mosabs} we can see
that  the   best  approximation  to  the  observed   DIG  surface  brightness
distribution is  found using an empirical coefficient  $k = 5.3\times10^{-23}
\times z_{100}$ cm$^{-2}$, and in Fig.~\ref{mosvarios} we can see the results of using this
value  with  two different  assumptions  about  the  Lyc escape  fraction:  a
constant escape fraction and  a constant fraction below $L_{H\alpha}=L_{Str}$,
and a  rising fraction for  higher \hii\ region  luminosities (cases A  and C
respectively) .  The results shown in  Figs.~\ref{mosabs} and \ref{mosvarios}
illustrate  the  fact  that  we  are  finding it  difficult  to  obtain  real
improvements in  our results with increasingly  refined assumptions.  Perhaps
Figs.~\ref{mosabs}b  y   \ref{mosvarios}b  between  them   give  the  closest
re\-pro\-duc\-tions of  the observations (e.g.  Fig.~\ref{mosabs}d),  and seem to
indicate that a  constant escape fraction for all the  regions gives the best
fit  to the  observed  DIG.  However  there  are a  few necessary  cautionary
statements to make.  We can see that Fig.~\ref{mosvarios}b, in which the fullest
set of assumptions has been brought in: Lyc field modified by absorption, and
emission in  \ha\ proportional  to the \hi\  column density, gives  a clearly
inferior reproduction of the DIG morphology than Fig.~\ref{mosabs}b, assuming
a constant \hi\ column density. This  is due to two effects, firstly the \hi\
column density  should not be used  in this way  where the \hii\ column  is a
major fraction  of the total H column,  as occurs near the  centre of NGC~157
(see Fig.~\ref{colden}  and discussion in  Sec.~\ref{discusion}). Secondly, we
are  coming up  against the  limits imposed  by the  relatively  poor angular
resolution in the  \hi\ map.  This is not so important  far from the sources,
where  the Lyc field  gradient is  low, but  is critical  close to  the \hii\
regions, and means  that models with any further  degree of refinement cannot
be properly empirically  tested. The combination of these  two effects can be
already leading  to misleading  conclusions in terms  of the fine  detail. An
apparent defect  in those  models where the  high luminosity regions  have an
increasing Lyc escape fraction is  that the predicted \ha\ near these regions
is  higher than  that observed.   Such  an effect  would be  expected if  the
\hii/\hi\ column density near the regions was raised (the same effect that we
see close to the centre of the galaxy in the constant \hi\ thickness model of
Fig.~\ref{mossinabs}a,c)  due to the  ionization itself. The  effect could be  detected only
with \hi\ resolution  of order 1\arcsec, i.e.  comparable to  that in \ha. In
the absence of such  data we are in no position to  detect the effect, so the
impression that the constant escape fraction models are better than the mixed
models could well be an incorrect interpretation.

\section{Discussion and conclusions}
\label{discusion}
We have used  a modelling technique to examine the  hypothesis that the \hii\
regions in disc galaxies are  sufficiently leaky that they allow the ionizing
photons  from their  OB stellar  populations  to escape  with intensity  high
enough to produce  the diffuse \ha\ observed across the disc  of a galaxy. We
have used  NGC~157 as a  test object here,  because one of  the observational
elements required in  the models is a  map of the \hi\ column  density of the
object to complement an existing map in calibrated \ha\ surface brightness. A
second element in  the modelling process is to  derive a catalogued intensity
calibrated map  of the individual \hii\  regions in \ha,  complemented by the
measured surface brightness  over the full diffuse component.  These data are
not trivial  to acquire, and  for the galaxies  for which we had  carried out
calibrated  \ha\ mapping  of the  necessary  precision, only  NGC~157 had  an
available map in \hi\ of anything approaching adequate angular resolution. In
spite  of the  approximations and  data limitations  mentioned freely  at the
relevant points  in the text, we  have found a very  fair coincidence between
the  distribution of  ionizing photons,  and the  
H$^{+}$   distribution   across   the   face   of  NGC~157,   as   shown   in
Figs.~\ref{mossinabs}, \ref{mosxh1}, \ref{mosabs}, and \ref{mosvarios}.  This $\,$
leads us to  conclude that in spite of the  simplicity and the approximations
with which the models have been ge\-ne\-ra\-ted, and the low angular re\-so\-lu\-tion of
the \hi\ data,  the hypothesis that the  ionization of the DIG in  NGC~157 is
caused by the Lyc photons escaping  from the \hii\ regions does receive major
support from this exercise.
	
The  fact  that  the  hypothetical   model  based  exclusively  on  the  \hi\
distribution,  and designed as  a test  of the  Sciama neutrino  decay theory
(Fig.~\ref{ha-h1}) gives  such complete  disagreement with the  observed \ha\
distribution in the DIG not only serves as a further rebuttal of this theory,
but  points up the rather  good agreement of the  models using leaky
\hii\ regions as the ionizing sources.

\begin{table}
\caption[]{Lyc continuum photon flux escaping from the \hii\ regions in NGC~157 
predicted by the models considered here, to be compared with the  Lyc flux (bottom
line) required to maintain the DIG in this galaxy ionized, assuming case B 
recombination and    T$\sim$10$^4$ K (Paper I).}
      \label{casos}
      \[
 \begin{array}{p{0.5\linewidth}l}
 \hline
  \hline
 {\bf Model/observation}& {\bf 10^{52}\,\mathrm{ Lyc\, \,phot\, \,s^{-1}}}\\
    \hline
Density  bounding (constant $\phi$)& 14.8\\
Density  bounding (constant $\rho$)& 29.4\\
Constant escape fraction 30\% & 6.81 \\
Mixed model& 15.8-30.4 \\
\hline
Required Lyc flux& 12-19\\
\noalign{\smallskip}
    \hline
    \hline
  \end{array}
      \]
\end{table}

        In order to estimate quantitatively the goodness of the models we
performed a series of numerical tests. These consisted of an analysis of the 
residuals obtained when subtracting the observed \ha\ emission of the DIG 
from the modelled emission (both observations and models normalized as 
explained, e.g. in Fig.~\ref{mossinabs}) via histograms showing the intensity distribution of 
all the pixels in these images of residuals. This method allowed us to 
discriminate between rather poor solutions (such as the Sciama model cited
above) or random \ha\  intensity distributions used as trials) and 
reasonably good solutions, which cover most of the cases in which the Lyc
photons escape from the \hii\ regions.

However, it is much more difficult using formal statistical tests to reach a
firmer conclusion about which of the assumptions about the escaping fraction 
of ionizing photons yields the best fit to the observations. The models which 
give low residuals are also those that seem to offer a better result by 
simple eye estimates, whereas those which give higher and more asymmetrically 
distributed residuals are also worse under visual inspection.

Direct inspection of the figures, and consideration of the corresponding
residual histograms give sufficient grounds to reject the
models  with the  highest  values of  the  empirical extinction  coefficient:
$k\gtrsim 5.3\times10^{-23}\times  z_{100}$ cm$^{-2}$, while  the model which
assumes that photons  escape only from the most  luminous regions, those with
$L_{H\alpha}>L_{Str}$, does not predict  sufficient intensity in the DIG near
the northernmost  arm of the  galaxy. Here the  majority of the  regions have
$L_{H\alpha}<L_{Str}$ but a measurable  surface brightness is measured in the
DIG surrounding the arm. It is  clear that flux must be escaping from regions
with $L_{H\alpha}<L_{Str}$,  a result found  also in models  where extinction
was  not considered. Very good fits  to the  observed DIG  surface brightness
distribution are found in those models  where we have taken a constant escape
fraction  for the Lyc  photons. This  appears to  go against  our hypothesis,
previously put  forward to  explain the form  of the \hii\  region luminosity
function in \ha\  (Beckman et al. 2000; Rozas et al.  1996a, 1999, 2000) that
\hii\  regions  with  $L_{H\alpha}>L_{Str}$  are  the  principal  sources  of
ionization of the  DIG.  However our models with  a constant escape fraction
of the ioni\-zing photons down--converted to \ha,
are normalized and yield maps with only
relative  intensity isophotes  in \ha.  To  fit the total  integrated
luminosity in the DIG of NGC~157  a constant escape fraction of 70\% would be
required  (see Table~\ref{casos}), and this  seems improbable from a range  
of physical considerations governing the escape of Lyc photons from star forming regions. 
Therefore it seems that if a constant fraction of ionizing photons is escaping from the \hii\ regions,
other principal ionizing sources (rather than \hii\ regions) 
would be needed in order to explain the power requirements to ionize the DIG 
(Table~\ref{casos}).

However, fits of very good quality are also found from models with a constant escape 
fraction of $\sim$20\% for low luminosity regions and an increasing fraction 
for those regions with $L_{H\alpha}>L_{Str}$, and this would be in better accord
with our deductions from luminosity functions and \hii\ region central surface 
brightnesses (Beckman et al. 2000).

We should note  here, that other ionizing sources may contribute to the ionization
of the DIG, and the aim of this paper is not to demonstrate that the \hii\  regions are the 
only ionizing source of the DIG. However, the results presented give strong support
to the hypothesis that  \hii\ regions are 
an important source and probably the main source for the ionization.
Known possible ionizing sources for the DIG in normal galaxies
include photoionization (by supernovae, Wolf-Rayets, white dwarfs, cosmic rays, X-rays, 
or field OB stars for example, Hoopes \& Walterbos 2000), but also shocks and turbulent 
dissipation in the ISM (e.g. Rand 2000; Walterbos 1999) may be important  in specific zones, 
as has been  noted within the Galaxy
 (Tufte, Reynolds \& Haffner 1999). 
Contributions from such sources to the general ionization of
the DIG are not excluded by the evidence presented here.

To go further,
reliably, into the  problem would require an effort  of detailed modelling of
transport in  a fully three--dimensional clumpy medium,  plus the acquisition
of  an  \hi\  map  of  sufficiently high  angular  resolution.  The  li\-mi\-ted
conclusion of  the present study  is that the  scenario in which  leaky \hii\
regions provide the ionizing sources  which yield at least the major fraction
of the \ha\ in the DIG appears to be well supported in the case of NGC~157.

\begin{acknowledgements}
This work was supported by the Spanish DGES (Direcci\'on General de 
Ense\~nanza Superior) via Grants PB91-0525, PB94-1107 and PB97-0219.  
The WHT is o\-pe\-ra\-ted on the island of La Palma by the Isaac Newton Group in
the Spanish Observatorio del Roque de los Muchachos of the Instituto de Astrof\'\i sica 
de  Canarias.
 
\end{acknowledgements}

\end{document}